\documentclass[twocolumn,prb,showpacs,superscriptaddress]{revtex4}

    \usepackage[dvips]{graphicx} 

\usepackage[english]{babel}
\usepackage{blindtext}
\usepackage{dcolumn}
\usepackage{bm}
\usepackage{ulem}
\usepackage[dvips]{color} 

\newcommand{\ie}{{\it i.e. }}

\newcommand{\co}[2]{\ifcase #1 \or #2 \fi}

\newcommand{\bscco}{Bi$_{2}$Sr$_{2}$CaCu$_{2}$O$_{8}$\,}

\newcommand{\celsius}{\,$^\circ$C}

\newif\ifnote

\begin{document}


\title{Interaction of hot spots and THz waves in Bi$_{2}$Sr$_{2}$CaCu$_{2}$O$_{8}$ intrinsic Josephson junction stacks of various geometry}

\author{S.~Gu\'{e}non}
\author{M.~Gr\"unzweig}
\affiliation{
Physikalisches Institut --
Experimentalphysik II
and
Center for Collective Quantum Phenomena,
Universit\"{a}t T\"{u}bingen,
Auf der Morgenstelle 14,
D-72076 T\"{u}bingen,
Germany
}
\author{B.~Gross}
\affiliation{
Physikalisches Institut --
Experimentalphysik II
and
Center for Collective Quantum Phenomena,
Universit\"{a}t T\"{u}bingen,
Auf der Morgenstelle 14,
D-72076 T\"{u}bingen,
Germany
}
\author{J.~Yuan}
\affiliation{National Institute for Materials Science, Tsukuba 3050047, Japan}
\author{Z.G.~Jiang}
\author{Y.Y.~Zhong}
\affiliation{Research Institute of Superconductor Electronics, Nanjing University, Nanjing 210093, China}
\author{A.~Iishi}
\affiliation{National Institute for Materials Science, Tsukuba 3050047, Japan}
\author{P.H.~Wu}
\affiliation{Research Institute of Superconductor Electronics, Nanjing University, Nanjing 210093, China}
\author{T.~Hatano}
\affiliation{National Institute for Materials Science, Tsukuba 3050047, Japan}
\author{D.~Koelle}
\affiliation{ Physikalisches Institut -- Experimentalphysik II and Center for Collective
Quantum Phenomena, Universit\"{a}t T\"{u}bingen, Auf der Morgenstelle 14, D-72076
T\"{u}bingen, Germany }
\author{H.B.~Wang}
\email{wang.huabing@nims.go.jp}
\affiliation{National Institute for Materials Science, Tsukuba 3050047, Japan}
\author{R.~Kleiner}
\email{kleiner@uni-tuebingen.de}
\affiliation{
Physikalisches Institut --
Experimentalphysik II
and
Center for Collective Quantum Phenomena,
Universit\"{a}t T\"{u}bingen,
Auf der Morgenstelle 14,
D-72076 T\"{u}bingen,
Germany
}%

\date{\today}

\begin{abstract}
At high enough input power in stacks of Bi$_{2}$Sr$_{2}$CaCu$_{2}$O$_{8}$ intrinsic Josephson junctions a hot spot (a region heated to above the superconducting transition temperature) coexists with regions still in the superconducting state. In the ``cold'' regions cavity resonances can occur, synchronizing the ac Josephson currents and giving rise to strong coherent THz emission. We investigate the interplay of hot spots and standing electromagnetic waves by 
low temperature scanning laser microscopy and THz emission measurements, using stacks of various geometries. For a rectangular and a arrow-shaped structure we show that the standing wave can be turned on and off in various regions of the stack structure, depending on the hot spot position. We also report on standing wave and hot spot formation in a disk shaped mesa structure.

\end{abstract}



\maketitle

\section{Introduction}
Recently, coherent THz emission with an extrapolated output power of some
$\mu$W from stacks (mesas) of more than 600 intrinsic Josephson junctions (IJJs) \cite{Kleiner92,Kleiner94a, Yurgens00} in \bscco (BSCCO) has been reported \cite{Ozyuzer07}, causing a lot of interest in using these structures as tunable generators 
\cite{Bulaevskii07, Koshelev08, Lin08,Koshelev08b, Hu08,Kadowaki08, Ozyuzer09a,Minami09,Kadowaki09,Kurter09,Gray09,Krasnov09,Lin09,Hu09,Klemm09,Nonomura09,Tachiki09,Koyama09,Savelev10,Wang09a, Wang10a, Yurgens10}. 
Phase synchronization involved a cavity resonance of the stack. The stacks used had much larger dimensions (in the 100 $\,\mu$m range) than the few-junction stacks investigated before. 
In those small sized structures, the IJJs in the stack tended to oscillate out-of-phase or were not synchronized at all
\cite{Hechtfischer97, Hechtfischer97b, Batov06, Bae07}.

While, in Refs.~\onlinecite{Ozyuzer07,Kadowaki08,Ozyuzer09a,Minami09,Kadowaki09,Kurter09,Gray09}, THz radiation from large stacks was reported for relatively modest dc input power where ohmic heating is not very severe, in Ref.~\onlinecite{Wang10a} coherent THz emission was reported at high bias. Here, as detected by low temperature scanning laser microscopy (LTSLM), a hot spot (\ie a region heated to above the critical temperature $T_c$) forms in the mesa coexisting with regions below $T_c$ \cite{Wang09a, Wang10a}. Standing waves can be imaged in the ``cold'' region. Hot spot and waves are correlated; particularly, the hot spot seems to adjust the size of the ``cold'' part of the mesa, making the cavity resonance frequency and thus the frequency of emission tunable. 
Given this role of the hot spot one can ask whether the hot spot can be manipulated in a controlled way to manipulate the THz radiation. Here we report on such experiments, showing by LTSLM that, in rectangular or arrow-shaped mesas, standing electromagnetic waves can be turned on and off by changing the position and/or the size of the hot spot.  
We also report on THz emission and hot spot and wave formation in a disk shaped mesa, as it has been proposed in Ref.~\onlinecite{Hu09}.

\section{Sample preparation and measurement techniques}
For the experiments BSCCO single crystals were grown by the floating zone technique in a four lamp arc-imaging
furnace. To provide good electrical contact the single crystals were cleaved
in vacuum and a 30\,nm Au layer was evaporated.
Then, conventional photolithography and Ar ion milling was used to prepare mesas of various shapes and thicknesses. 
Insulating polyimide was used to surround the mesa edges at which Au wires were attached to the mesa by
silver paste. Other Au wires were connected to the big single crystal pedestal as grounds.
Below we discuss results from four samples. 
The arrow shaped sample 1, c.f. Fig.\ref{fig:1}, consisted of three 330 $\mu$m long and 50 $\mu$m wide rectangular subsections rotated by $60^\circ$ relative to each other and connected at one end (the top of the arrow).  The mesa thickness was  1 $\mu$m, corresponding to 670 IJJs. It was patterned on a crystal that was annealed in vacuum at 600\celsius\ for 72 hours. The crystal had a $T_c$ of 87.6 K and a transition width $\Delta T_c$ of 1.5 K. The mesa was contacted at the end of the main stem of the arrow. 
The rectangular sample 2 was 330 $\mu$m long, 80 $\mu$m wide and 1 $\mu$m thick. It was  patterned on a crystal ($T_c$ = 83 K, $\Delta T_c$ = 1.5 K) annealed in 1 atm (Ar 99$\%$+O$_2$ 1$\%$ ) at 600\celsius\ for 48 hours. The mesa was contacted at two edges, c.f. Fig.\ref{fig:2} (a).
The $330\times 70$ $\mu$m$^2$ large  and 0.5 $\mu$m thick rectangular sample 3, contacted on the right edge, c. f. Fig.\ref{fig:3}(b), was surrounded by much smaller, 10 $\mu$m wide, quadratic mesa structures consisting of 15--20 IJJs. The distance of these mesas (used for thermometry) to the main mesa was 10 $\mu$m. This sample was patterned on a crystal ($T_c$ = 87.6 K, $\Delta T_c$ = 1.5 K) annealed in vacuum at 600\celsius\ for 72 hours. 
Sample 4 was disk shaped, with a radius of 100 $\mu$m and a thickness of 0.9 $\mu$m. It was  patterned on a crystal ($T_c$ = 86.6 K, $\Delta T_c$ = 1.5 K) annealed in vacuum at 650\celsius\ for 65 hours. The mesa was contacted at near its outer edge, as sketched in Fig.\ref{fig:4} (a). 

In order to provide a load line for stable operation, the mesas were
biased using a current source and variable resistor in parallel to
the mesa, cf.~Fig. 1 in Ref.~\onlinecite{Wang09a}.
The voltage measured across the mesa includes the resistance of the contacting Au wires and the resistance between these wires and the mesa.  In the data discussed below this resistance (typically around 5$\Omega$) is subtracted.

LTSLM measurements were performed in T\"ubingen. For sample 4, additional THz emission measurements were performed in Tsukuba. 

The LTSLM setup is described in Ref.~\onlinecite{Wang09a}. In brief, the beam of a diode laser (modulated at 10-80 kHz, spot size 1-2 $\,\mu$m) is deflected by a scanning unit and focused onto the sample surface. Local heating by 2--3 K in an area of a few $\mu$m$^2$ and about 0.5 $\mu$m in depth causes a change $\Delta V$ (detected using lock-in techniques) of the voltage $V$ across the mesa  serving as the contrast for the LTSLM image. Primarily, the laser beam causes changes in both the in-plane and the out-of-plane 
critical supercurrent densities and resistivities. 

Standing waves can be imaged due to the beam-induced local change of the quality factor, leading to a strong signal $\Delta V$ at antinodes and a weak signal at nodes. Whether dominantly the (squares of the) electric or magnetic field components are imaged may depend on both the bias and the bath temperature. For Nb tunnel junctions the response was shown to primarily arise from the magnetic field component \cite{Lachenmann93}. For IJJ stacks the situation is less clear; however, below we will show an example (sample 4) where the LTSLM response was due to the magnetic field.

A typical signature of the hot spot is that it grows with increasing input power. Its edge leads to a strong (negative) LTSLM signal, while $\Delta V$ is small in the interior of the hot spot and in the ``cold'' part of the mesa \cite{Wang09a}. 

The THz emission setup is described in detail in Ref.~\onlinecite{Wang10a}. In brief, the crystal is mounted in a continuous flow cryostat with a polyethylene window. The emitted radiation is collected by a parabolic mirror, deflected towards two lamellar split mirrors (one of which is mounted on a translation stage) and finally guided to a bolometer via a second parabolic mirror. A 1 THz cut-off type filter is used in front of the bolometer. Either the bolometric response was recorded directly or frequency spectra were taken via Fourier transform of the autocorrelation function when moving one of the split mirrors. The solid angle of the setup, defined by the aperture of the Winston cone in front of the bolometer, is 0.04 sr. 

\section{Results}


\begin{figure}[tb]
\begin{center}
\includegraphics[width=\columnwidth,clip]{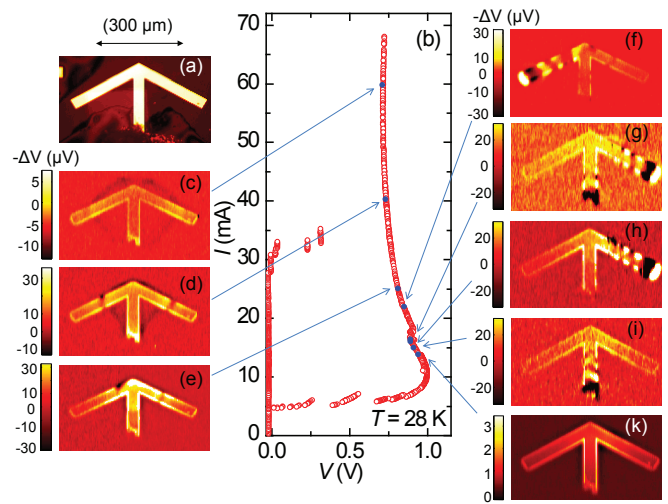}
\end{center}
\caption{(color online) Optical image (a), current voltage characteristic (b) and LTSLM images (c) to (k) of the arrow shaped sample 1, as measured at T = 28 K. 
}
\label{fig:1}
\end{figure}
%

In previous experiments \cite{Wang09a,Wang10a} we have seen that, using rectangular mesas, the hot spots tend to form near the contacting wire. Figure \ref{fig:1} (a) shows the more complex structure of an arrow shaped mesa (sample 1) that can be viewed as three rectangular parts connected at one end. The current voltage characteristic (IVC), taken at a bath temperature of 28 K, is shown in Figure \ref{fig:1}(b). Starting from zero current the IJJs in the mesa switch to their resistive states at currents $I$ between 30 mA and 35 mA. When lowering the current from high bias one observes an instability for currents  between 18 mA and 16 mA that is indicative for the disappearance of a hot spot present at high bias. LTSLM images are shown in Figures  \ref{fig:1} (c) to (i), taken in the resistive state of all junctions.  For currents decreasing from 60 mA to 25 mA one observes a ring-like feature centered around the top of the arrow and decreasing size with decreasing input power, c.f. Figures \ref{fig:1} (c), (d) and(e), the inside of which forms the hot spot. Such a feature has been seen for several rectangular mesas \cite{Wang09a,Wang10a}. Typically, there is no response from \textit{outside} of a mesa, in contrast to sample 1, where the hot spot can be seen even far away from the mesa edges. However, it can happen that at the foot of the mesa, after ion milling, there are a few weakened CuO$_2$ layers which contribute to the voltage measured, making also structures visible that are outside the actual mesa. Figures \ref{fig:1} (c), (d) and(e) show that, at high input power, the hot spot  is not restricted to the actual mesa but extends far beyond it (also note the dark feature at the foot of the arrow, which is the wire contacting the mesa). When lowering the current, between 24 mA and 15 mA standing wave patterns appear in various parts of the mesa.
Near $I=$22 mA the cavity formed by the left part of the arrow is excited [Fig. \ref{fig:1}(f)]. At $I=$16.4 mA a wave is visible in the right part of the arrow and the shaft [Fig. \ref{fig:1}(g)], at $I=$16.0 mA the wave is visible only in the right part of the arrow [Fig. \ref{fig:1}(h)]  and at $I=$15.0 mA the wave appears only in  the shaft [Fig. \ref{fig:1}(i)]. For currents below 15 mA the LTSLM images were smooth and neither hot spot nor wave features appeared, c.f. Fig. \ref{fig:1}(k).
Thus, in a narrow current range the arrow acted as a switch, generating standing waves in different arms of the structure. This property can be understood if one assumes that the hot spot separates the arrow effectively into three cavities which can become resonant or not, depending on the position of the hot spot edges adjusting the size of the cavities. In Figure \ref{fig:1}(f) the hot spot is clearly visible. For Figures \ref{fig:1}(g), (h) and (i) we cannot unambiguously tell. However, these images have been taken very close to the instability in the IVC (kink at 18 mA), making it likely that there is either still a small hot spot or at least a zone very close to $T_c$.
Standing waves appeared at voltages between 0.8 V and 0.9 V which, according to the Josephson relation, for this 670 IJJ stack would correspond to oscillation frequencies between 0.55 and 0.58 THz (we have not measured emission explicitely for this sample). From the standing wave patterns we infer wavelengths around 100$\,\mu$m and from that mode velocities $c'=f\lambda$ of $(4-6)\cdot 10^7 m/s$, which is compatible with the mode velocity expected for all junctions oscillating in-phase \cite{Wang09a,Wang10a}. The oscillation and wave properties of the arrow are thus comparable to our observations for rectangular mesas \cite{Wang09a,Wang10a}.

\begin{figure}[tb]
\includegraphics[width=\columnwidth,clip]{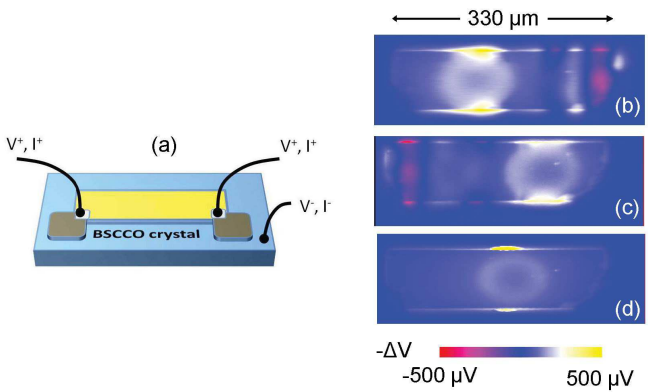}
\caption{(color online). 
Sketch of sample 2 (a) and LTSLM data at $T=$20 K for various biasing conditions: (b) bias from left, (c) bias from right, (d) symmetric bias.
}
\label{fig:2}
\end{figure}
%

The data of sample 1 suggest that cavity resonances can be turned on and off by changing the size and position of the hot spot. Thus, it should be possible to create a similar effect in a more controlled way by \textit{moving} a hot spot  within a given cavity at fixed bias current. To test this we contacted a rectangular $330\times 80$ $\mu$m$^2$ mesa (sample 2) at two edges, cf. Figure \ref{fig:2}(a). In Figures  \ref{fig:2}(b), (c) and (d) we show LTSLM images taken at $T$=40 K at a bias current of 30.4 mA. For image  \ref{fig:2}(b) the current has been injected at the left contact. Here, the hot spot appears as the circular structure in the left half of the mesa and there is a clear standing wave pattern in the right half of the stack. For the image \ref{fig:2}(c) the current has been injected from the right side. The hot spot now appears in the right half of the mesa and the wave is situated in the left part. Finally, Fig. \ref{fig:2}(d) shows a symmetric situation where the current has been injected from both edges. Here, the hot spot is centered in the mesa and no wave appears. For these measurements we have moved the hot spot at fixed bias current. The voltage across the sample (and thus the Josephson frequency) slightly varied for the three bias conditions, from 0.589 V when biasing from left, via 0.577 V when biasing from the right, to 0.619 V for the symmetric bias. To check whether this change of voltage is the dominant quantity for the (dis)appearance of the standing wave we have investigated hot spot and wave formation for the symmetric bias and the bias-from-left configuration over a wide current range. It turned out that, for the symmetric bias, we could not achieve a wave pattern at all. For the other configuration the wave appeared over a wide current range, from about 28 mA to 45 mA. Here, the voltage varied between 0.51 V and 0.63 V, i.e. over a much larger range than for the case of Figures  \ref{fig:2}(b), (c) and (d).

\begin{figure}
\includegraphics[width=\columnwidth,clip]{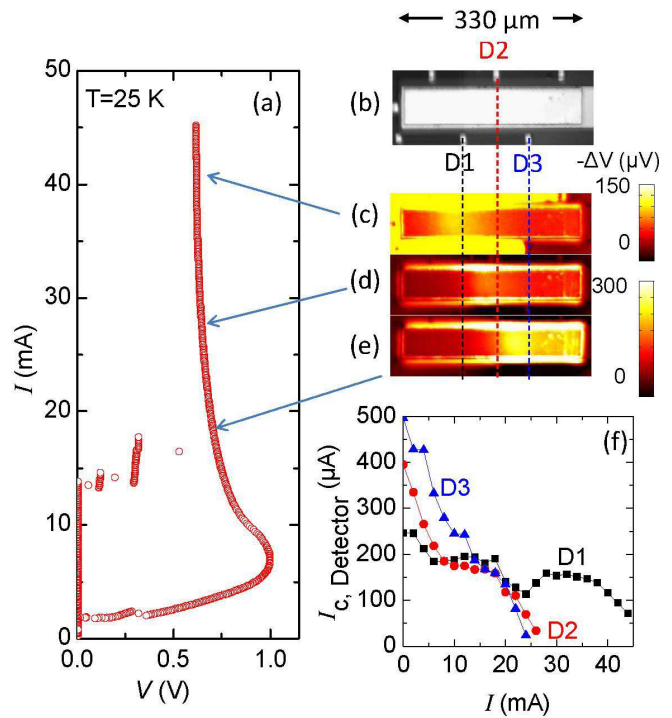}
\caption{(color online)
Current voltage characteristic at $T$=25 K (a) of sample 3 (main mesa) together with optical image (b), LTSLM images (c),(d),(e) and the critical current of detector junctions D1,D2 and D3 vs. bias current through main mesa (f).  In (c),(d) and(e) the bias current through the main mesa is, respectively, 42.4 mA, 28.8 mA and 18.6 mA. 
} 
\label{fig:3}
\end{figure}

The LTSLM data of the arrow shaped sample 1 further suggested that the hot spot extends well beyond the actual mesa structure. To test this further we have fabricated a rectangular $330\times 70$ $\mu$m$^2$ large  and 0.5 $\mu$m thick sample 3 which was surrounded by small mesas acting as thermometers. These mesas were separated by  10 $\mu$m  from the main mesa. Three of them (denoted D1, D2 and D3) worked. An optical image of the main mesa and the surrounding detector mesas is shown in Figure \ref{fig:3}(b). The working detector mesas are indicated. The current $I$ through the main mesa is injected from its right side. Figure \ref{fig:3}(a) shows the IVC of the main mesa, as measured at $T$=25 K.  At $I<$ 10 mA a hot spot forms near the right edge of the mesa. With increasing bias its left edge moves to the left, as shown in the LTSLM images \ref{fig:3}(c), (d) and (e). Fig. \ref{fig:3}(f) shows the critical current $I_c$ of the three detector junctions as a function of the current through the main mesa.  As can be seen from \ref{fig:3}(e), for $I$=18.6 mA the hot spot edge (at least as seen inside the main mesa) has just passed  detector D3. $I_c$ of this detector goes to zero at the not much higher current $I \approx$ 22 mA, where $I_c$ of detectors D1 and D2 are still finite. At $I$=26.8 mA the front of the hot spot has passed detector D2, c.f. Figure \ref{fig:3}(d), consistent with the observation that  $I_c$ of this detector vanishes at $I \approx$ 25 mA. Finally, $I_c$ of detector D1 approaches zero near $I=$ 45 mA. For comparison, the front of the hot spot passes this detector for $I\approx$43 mA, c. f. Figure \ref{fig:3}(c). The fact that the critical currents of the detector junctions vanish when being passed by the front of the hot spot, on one hand demonstrate that our hot spot interpretation of the corresponding feature observed in LTSLM is correct and, on the other hand, shows that the hot spot indeed is not restricted to the mesa itself. We further note that, for sample 3, we could not detect standing wave features.

\begin{figure}[tb]
\includegraphics[width=\columnwidth,clip]{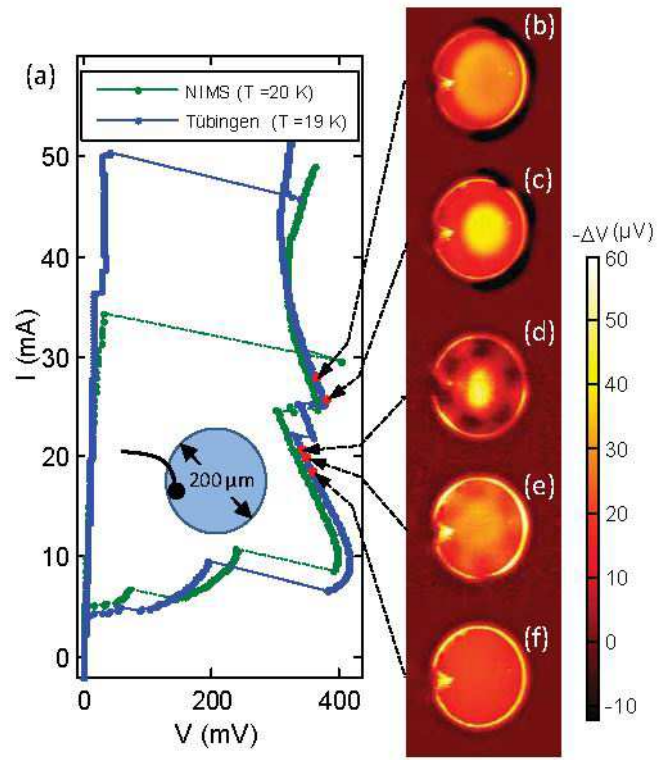}
\caption{(color online). IVCs, as measured in T\"ubingen (solid line and blue circles) and in Tsukuba (dashed line and green circles) (a) and LTSLM images (b) to (f) of disk shaped sample 4. Bias points (refering to the T\"ubingen IVC) where the images have been taken are indicated by arrows. Inset in (a) shows a sketch of the disk including the contacting lead.
}
\label{fig:4}
\end{figure}
%

%
\begin{figure}
\includegraphics[width=\columnwidth,clip]{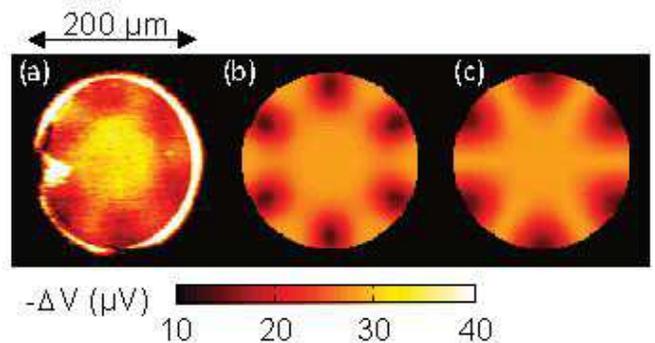}
\caption{(color online)
LTSLM image (a), as measured at  $I = 19.4$ mA in comparison of calculated squares of magnetic field (b) and electric field (c) of the $m=3$ $n=1$ cavity mode.
} 
\label{fig:5}
\end{figure}

\begin{figure}
\includegraphics[width=0.8\columnwidth,clip]{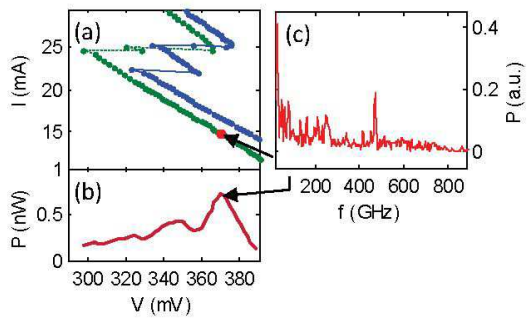}
\caption{(color online)
THz emission from sample 4: (a) enlargement of the IVC of Fig. \ref{fig:5}(a) in the current and voltage range where emission has been detected; (b) detected emission power vs. voltage across mesa and (c) Fourier spectrum at a bias of 14.9 mA and 370 mV. 
} 
\label{fig:6}
\end{figure}

We now return to nonrectangular geometries and discuss data of disk-shaped sample 4 which had a radius $a$ of 100 $\mu$m. Disk shapes mesa structures have been analyzed and considered promising for THz emission in Ref.~\onlinecite{Hu09}. In our context we are, on one hand, interested in imaging the corresponding wave patterns. On the other hand the question arises whether or not hot spot formation is supporting cavity modes in this geometry. 
In the disks, the c axis electric field component is given by the time derivative of a term $\tilde{P}$ describing Josephson plasma oscillations at frequency $\omega$ and having the form $\tilde{P}(r,t)=Ag_{mn}(r)\sin(\omega t+\varphi)$, with some amplitude $A$ and some initial phase $\varphi$. In cylindrical coordinates $(\rho,\phi)$ the function $g_{mn}(r) = J_m(\chi^c_{mn}\rho/a)\cos(m\phi)$ where $\chi^c_{mn}$ is the $n$th zero of the derivative of Bessel function $J_m$. The in-plane magnetic field is proportional to the curl of $\tilde{P}\cdot e_z$, with the out-of-plane unit vector $e_z$. The resonance frequency is given by $c'\chi^c_{mn}/2\pi a$, with the mode velocity $c'$.    
Fig. \ref{fig:4} (a) shows by a solid line and  blue circles the IVC of sample 4, as measured in T\"ubingen at $T=$19 K (the IVC shown by a dashed line and green circles has been measured in Tsukuba and will be discussed below in the context of THz emission measurements). For bias currents between 22 mA and 25 mA there are instabilities on the resistive branch indicative of hot spot formation \cite{Wang09a}. In addition, some of the junctions may switch between the resistive state and the zero voltage state (indeed one can trace out several different branches in this region by sweeping the bias current back and forth). Figures \ref{fig:4} (b) to (f) show LTSLM images taken at different bias points. The bias lead, attached on the left side of the disk, as sketched in the inset of Fig. \ref{fig:4} (a), is visible in all LTSLM images. Apart from this signal and edge effects (the mesas edges are not covered by gold and the laser beam induced temperature rise is larger), for currents below 20 mA  the images are smooth, c.f. Fig.  \ref{fig:4} (f). With further increase of current a standing wave pattern appears (at $I =$ 19.4 mA) having 6 nodes in azimuthal direction. In Fig. \ref{fig:4} (d), in addition a hot spot has formed in the center of the disk which is not visible in the image of  Fig.  \ref{fig:4} (e). 
With increasing current this hot spot grows, c.f. Figures \ref{fig:4} (b) and (c) but no more wave features appear. 
The standing wave pattern seen is indicative of an $m=3$, $n=1$ mode. Here, $\chi^c_ {mn}$=4.2. In LTSLM the signal $\Delta V$ is proportional to either the square of electric or magnetic field. While for rectangular geometries, due to edge effects, it is very difficult to decide which component matters, the observation of the standing wave in the disk now allows to address this question.
Fig. \ref{fig:5} compares the standing wave pattern (a), measured at $I = 19.4$ mA, to the calculated squares of magnetic field (b) and electric field (c). As can be seen, image (b) reproduces the hexagon shaped center part of the LTSLM image much better than image (c) where the center part is more star shaped. This strongly indicates that the LTSLM signal is dominantly sensitive to magnetic field variations. In Fig.  \ref{fig:6} 
we analyze the emission detected (in Tsukuba) from this sample.  Fig.  \ref{fig:6} (a) shows an enlargement of the IVC of Fig.  \ref{fig:4} (a) for the current and voltage range where emission has been detected (the same range where the cavity resonance was seen in LTSLM; note however, that the IVCs measured in T\"ubingen and in Tsukuba slightly differ from each other, making it difficult to precisely assign LTSLM images and emission data). Emission peaks appear in the voltage range 330--380 mV, c.f. Fig.  \ref{fig:6} (b), with a maximum \textit{detected} power of about 0.8 nW. Extrapolated over $4\pi$ this yields a total power of order 0.25 $\mu$W which is actually almost an order of magnitude below the radiation power detected for rectangular stacks \cite{Ozyuzer07, Wang10a}. 
Fig. \ref{fig:6} (c) shows a Fourier spectrum of this radiation, with the sample biased at 14.9 mA and 370 mV (corresponding to LTSLM image Fig. \ref{fig:4}(e) where no hot spot has formed yet). The radiation peak occurs at $f=$472 GHz. Using the Josephson relation,  $f=NV/\Phi_0$, with the flux quantum $\Phi_0$, we infer a junction number $N=380$, which only about 60$\%$ of the 600 junctions contained in the mesa, as estimated from its 0.9 $\mu$m thickness. Thus, not all junctions participate in radiation. Finally, from the $m=3$, $n=1$ mode observed and from the emission frequency we find a mode velocity of $c'\approx 7\cdot10^7$ m/s, which is the value to be expected when the junctions oscillate in-phase and the temperature is not too high\cite{Wang09a}. 
At higher bias we have seen from Fig. \ref{fig:4}(d) that a hot spot forms in the center of the mesa. The resonance still corresponds to a $m=3$ mode. Then, the mesa effectively might be considered as an \textit{annular} mesa, with an inner radius $a_i$ corresponding to the hot spot radius.  The resonance frequency is given by  $c'\chi_{mn}/2\pi a$, where the coefficient $\chi_{mn}$ depends on the ratio $a_i/a$, c.f. equation (15) in Ref.~\onlinecite{Hu09}. Depending on $n$ and $m$ it can either increase or decrease as a function of $a_i/a$. For the $m=3$, $n=1$ mode, $\chi_{mn}$ is almost constant for $a_i/a<0.3$.  For the case of Fig. \ref{fig:4}(d) the size of the hot spot is still small, $a_i/a<<1$, and $\chi_{mn} \approx \chi^c_{mn}$. On the other hand, with increasing temperature in the mesa the mode velocity $c'$ decreases \cite{Wang09a} and thus effectively lowers the resonance frequency. Thus, as for rectangular mesas, also resonance modes of disk shaped mesas can be tuned to some extent by hot spots. However, the tunability seems to be more restricted than for rectangular geometries. At least for the sample discussed there was no more resonance as soon as the hot spot filled a significant fraction of the stack, cf. Figs. \ref{fig:4} (b) and (c). 


\section{Conclusions}
In conclusion, we have investigated hot spot and wave formation in a variety of different mesa geometries. To excite a cavity mode in the mesa, the frequency of the Josephson currents, determined by the voltage across each intrinsic junction which in turn is determined by the effective mesa temperature must match the cavity resonance frequency. While the resonance condition for a homogeneous structure is hard to fulfill in general, the appearance of a hot spot makes the cavity modes tunable. 
In rectangular mesas the hot spot often nucleates in the vicinity of the current injection point, leaving part of the mesa below the transition temperature $T_c$. The size and also temperature of this ``cold`` part is adjusted by the hot spot position. Its resonance frequency can be tuned via the cavity size and the temperature dependent mode velocity. 
Indeed, we have turned on and off standing waves by moving the hot spot using two different current injection points. In an arrow shaped geometry the hot spot nucleated at the tip of the arrow, dividing this structure effectively into three rectangular subsections. In each of them standing waves could be excited, depending on the hot spot size and position. We also have seen that the hot spot extends well beyond the actual mesa, possibly allowing to thermally couple to different mesas located not too far from each other. Hot spots and waves also appeared in a disk shaped mesa. The hot spot, formed in the center of the mesa, effectively turns the disk into a ring shaped structure. These results show that controlled hot spot formation is a powerful tool to manipulate and tune standing waves and thus THz emission from intrinsic junciton stacks. It should be taken into account when designing future tunable THz generators based on intrinsic Josephson junction stacks. 

\section{Acknowledgments}
We thank  A. Yurgens, V.M.Krasnov, U. Welp, L.Ozyuzer, K. Kadowaki, I. Iguchi, K. Nakajima and C. Otani for valuable discussions. 
Financial support by the strategic Japanese-German International Cooperative Program of
the JST and the DFG, and Grants-in-Aid for scientific research from JSPS is gratefully acknowledged. 


%
%
\bibliography{./bib/hot-spots-waves_v2}
%
\end{document}